# Image-to-Image Translation with Generative Adversarial Network for Electrical Resistance Tomography Reconstruction

Weijian yan

## Abstract

Electrical tomography techniques have been widely employed for multiphase-flow monitoring owing to their non-invasive nature, intrinsic safety, and low cost. Nevertheless, conventional reconstructions struggle to capture fine details, which hampers broader adoption. Motivated by recent advances in deep learning, this study introduces a Pix2Pix generative adversarial network (GAN) to enhance image reconstruction in electrical capacitance tomography (ECT). Comprehensive simulated and experimental databases were established and multiple baseline reconstruction algorithms were implemented. The proposed GAN demonstrably improves quantitative metrics such as SSIM, PSNR, and PMSE, while qualitatively producing high-resolution images with sharp boundaries that are no longer constrained by mesh discretisation.

## 1. Introduction

Electrical capacitance tomography (ECT) reconstructs the permittivity distribution inside a sensing domain from boundary measurements acquired by an electrode array. Owing to its "soft-field" nature and the limited number of electrodes, conventional reconstructions often suffer from low resolution. A plethora of algorithms—e.g., Landweber [1], Newton–Raphson [2], Tikhonov regularisation [3], and conjugate-gradient (CG) methods [4]—have been proposed to mitigate this issue. More recently, deep-learning-based approaches have emerged. Li *et al.* (2018) [5] pioneered the use of convolutional neural networks (CNNs) for ERT; Fabijańska *et al.* (2020) [6] employed graph neural networks for 3-D ECT, and Wu *et al.* (2021) [7] adopted restricted Boltzmann machines. Among deep models, GANs are particularly attractive for unsupervised learning on complex data distributions. A GAN comprises a *generator* that synthesises images and a *discriminator* that evaluates authenticity; the two networks are trained adversarially.

This paper proposes an ECT reconstruction algorithm based on the conditional Pix2Pix-GAN [8]. A U-Net serves as the generator, whereas a patch-based discriminator evaluates local realism. Three classical algorithms—LBP [9], Landweber, and Tikhonov—were used to create 6 000 grayscale reconstructions from simulation and experimental datasets. Thirty percent of these images formed an independent test set. The GAN-enhanced images exhibit significant gains in SSIM, PSNR, and PMSE and, more importantly, display arbitrarily shaped objects with crisp contours, overcoming the mesh-grid artefacts of traditional methods. Owing to Pix2Pix's image-to-image nature, the method is agnostic to electrode number and mesh topology; this was corroborated experimentally under 8- and 32-electrode configurations.

## 2. Methodology

### 2.1 ERT Reconstruction

As shown in Fig. 1, image reconstruction in ERT [10] infers the spatial distribution of the medium within the domain from boundary current measurements via suitable algorithms. ERT is modeled as a steady (constant-current) field, in which the current density vanishes everywhere in the domain. According to Maxwell's equations, the mathematical model of the ERT sensor can be expressed as:

$$\begin{cases} \nabla \cdot (\sigma \cdot \nabla \varphi) = 0 \text{ within domian} \\ \sigma \frac{\partial \varphi}{\partial n}\Big|_{\Gamma_{E+}} = +j \text{ input current} \\ \sigma \frac{\partial \varphi}{\partial n}\Big|_{\Gamma_{E-}} = -j \text{ output current} \\ \sigma \frac{\partial \varphi}{\partial n}\Big|_{\Gamma_{E}} = 0 \text{ ouside domian} \end{cases} \#(1)$$

where $\nabla \cdot$ and $\nabla$ denote the divergence and gradient operators, respectively; σ is the electrical conductivity distribution of the material within the pipe cross-section; $\Gamma_{E+}$ and $\Gamma_{E-}$ are the electrode boundaries at which current enters and leaves the domain, respectively; $\Gamma_E$ is the boundary with no current exchange (an insulating/no-flux boundary); and j is the line current density on the driving electrode, equal to the excitation current divided by the electrode width.

After discretizing and normalizing the nonlinear relationship between the inter-electrode capacitances and the permittivity distribution, the resulting linearized equation can be written approximately as follows:

$$G = US^{-1} \#(2)$$

Here, S denotes the sensitivity matrix; G is the permittivity vector (i.e., the gray-level image vector); and U is the measured data. The problem thus reduces to estimating G from the known U. However, owing to the soft-field nature of ERT, the sensitivity field does not faithfully capture the mapping from the medium distribution to the measured capacitances, which often leads to suboptimal image quality. In this work, we reconstruct images from the raw data using the LBP, Landweber, and Tikhonov algorithms. The essence of LBP is to approximate $S^{-1}$ by $S^T$, i.e., to solve for G using $S^T$ in place of $S^{-1}$.

$$G = US^T \#(3)$$

The essence of the Landweber algorithm is to compute the generalized inverse of the sensitivity matrix via an iterative method. Its expression is:

$$G_{k+1} = G_k + \alpha_k S^T(V - SG_k) \#(4)$$

where $G_k$ is the gray-level image vector at the k-th iteration; S is the sensitivity matrix; V denotes the boundary-voltage measurements; and $\alpha_k$ is the gain (step-size) factor, satisfying …

$$\|\alpha_k S^T S\|_2 < 2 \#(5)$$

Tikhonov regularization is a commonly used non-iterative reconstruction method; its canonical regularized solution is given by:

$$G = P(S^T S + \lambda I)^{-1} S^T V \quad \#(6)$$

where λ is the regularization parameter, I is the identity matrix, and P is a 0–1 (binary) operator.

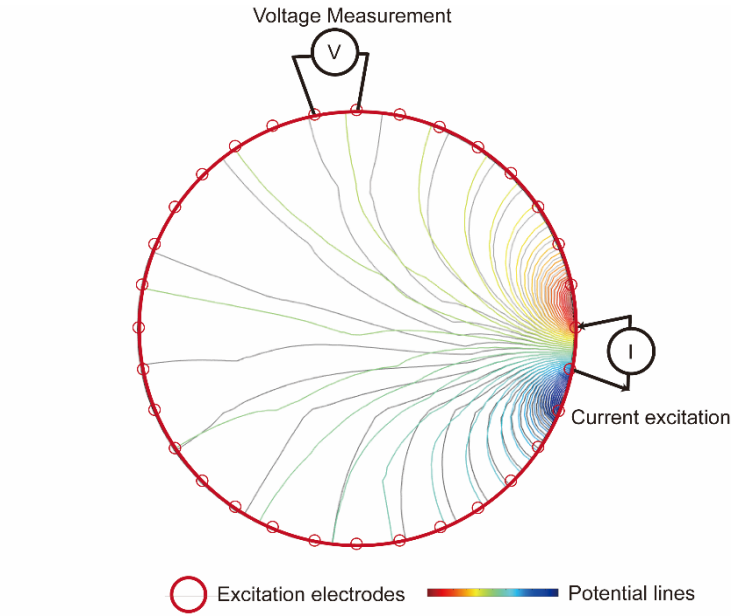

Fig. 1. 32-electrode ERT model

## 2.2 Structure of Pix2pixGAN for ERT Reconstruction

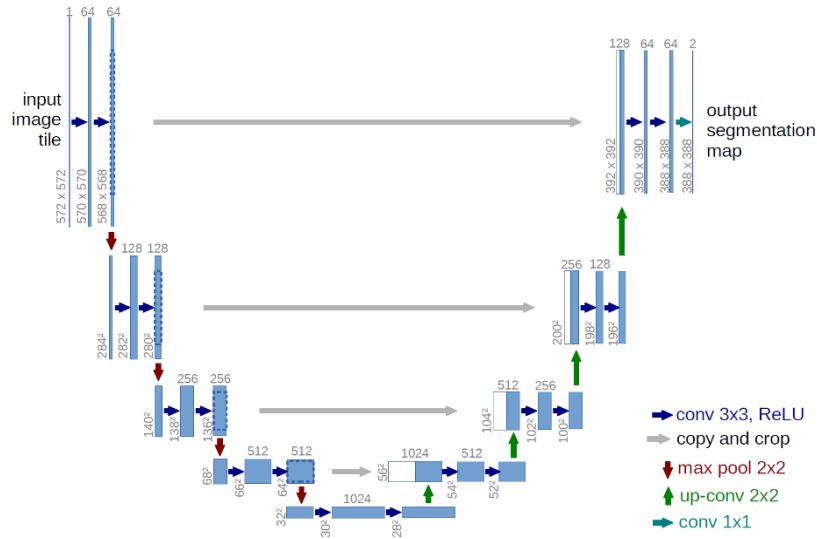

Fig. 2. U-net structure[13]

The network used in this study comprises a generator and a discriminator. The generator adopts a U-Net [11] architecture, and the discriminator follows the PatchGAN [12] design. U-Net is a fully convolutional neural network for image segmentation; its architecture is shown in Fig. 2. It has two key features: skip connections and transposed-convolution (deconvolution) layers. Skip connections link the encoder and decoder so that the decoder can recover fine-grained details captured by the encoder, alleviating "detail loss" in segmentation. The transposed-convolution layers upsample feature maps in the decoder to restore spatial resolution, enabling high-resolution segmentation outputs.

In the original GAN, the discriminator outputs a single scalar score that evaluates the entire generated image. By contrast, PatchGAN is designed as a fully convolutional discriminator: after a series of convolutions, the network does not feed into a fully connected head; instead, a final convolution maps the input to an N×N matrix of scores, serving the role of the scalar realism score in the vanilla GAN. Each element of this matrix corresponds to a small patch of the input image and evaluates the realism of that local region, as illustrated in the figure 3.

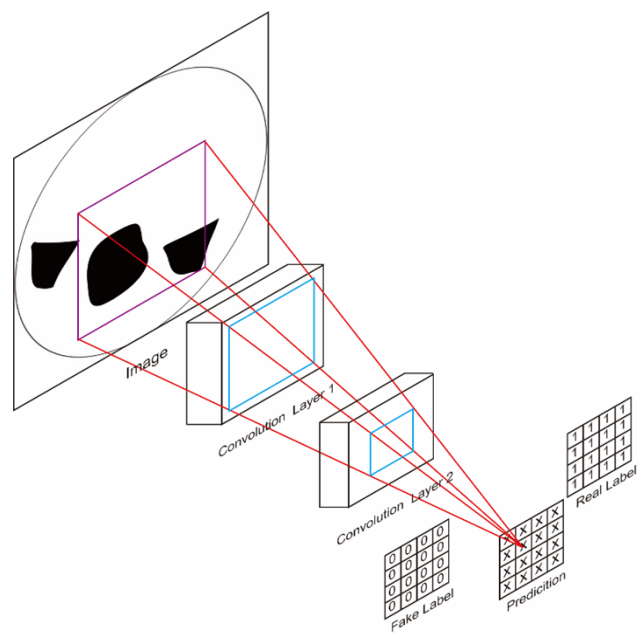

Fig. 3. Patch-based discriminator

As illustrated in Fig. 4, data acquired from experimental instruments or numerical simulations are first reconstructed by the ERT imaging algorithm to produce a grayscale image, which is then fed into the generator network. The generator produces an image based on this input and passes it to the discriminator. In parallel, the corresponding real image is also provided to the discriminator, which compares the two, judges the authenticity of the generated image, and feeds back a learning signal to the generator. Through this adversarial process, the image quality is iteratively improved until the discriminator classifies the generated images as real.

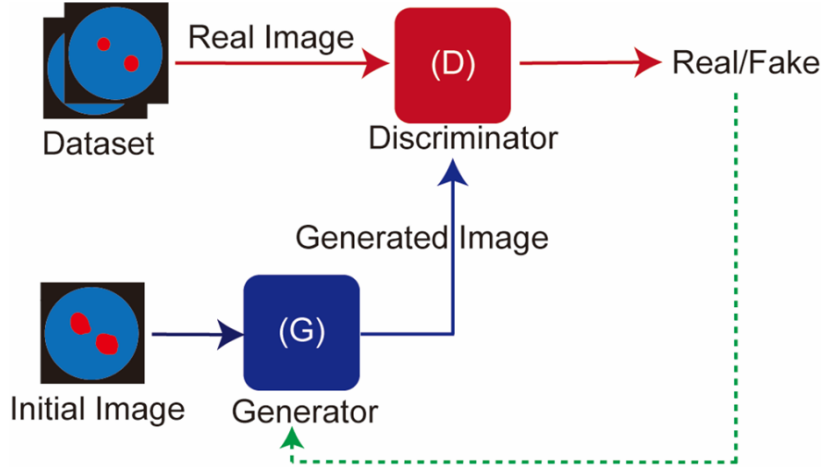

Fig. 4. Pix2pix GAN structure for ERT reconstruction

## 2.3 Objective

The objective of pix2pix can be written as

$$L_{cGAN}(G,D) = E_{x,y}[log D(x,y)] + E_{x,z}[\log(1 - D(x,z))]$$

Here, G attempts to maximize the objective, while D attempts to minimize it, which can be expressed as:

$$G^* = argmin_G max_D L_{cGAN}(G,D)$$

By comparing this objective with the original formulation, it is observed that blending the generated image with the ground truth image is beneficial. To further encourage the output to be close to the real image, an L1 distance is used instead of L2, since L1 tends to reduce blurring. The modified term is

$$L_{l1} = E_{x,y,z}[\|y - G(x,z)\|_1]$$

The resulting loss function is:

$$G^* = argmin_G max_D L_{cGAN}(G,D) + \lambda L_{l1}(G)$$

## 3. Stimulation & Experiment

To provide the training and test datasets required by the neural network, we performed simulations in MATLAB using the finite-element method (FEM) with second-type (Neumann) boundary conditions. As shown in Fig. 5, the ERT simulation domain was uniformly triangulated into 1,024 elements. Either 32 or 8 electrodes were uniformly distributed along the circular boundary. To emulate different material distributions, a random number of circular, triangular, and square inclusions were generated at random locations. The conductivities of the simulated inclusions were assigned prescribed values, while the background (air) was set to zero conductivity. The excitation voltage on the boundary electrodes was 5 V. In total, 6,000 datasets were generated.

For experiments, we employed ERT sensors with 8 and 32 electrodes (Fig. 6). The insulating tube had an inner diameter of 10 cm and an outer diameter of 12 cm, and each sensing electrode had an axial length of 38 mm. The relative permittivity of the outer insulating pipe was 4, and that of the shielding layer was 2.5. The test material consisted of quartz granules with a relative permittivity of 2.2. Conductivity was measured using an AC method. A quartz rod was inserted into the ERT sensor; data were acquired by the DAQ system and transmitted to a PC via Ethernet.To verify practical applicability, a quartz rod (conductivity 2.5) was inserted into the ECT pipeline to simulate a target medium distribution.

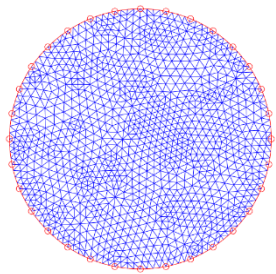

Figure. 5. Electrode layout and mesh discretization in simulations and experiments

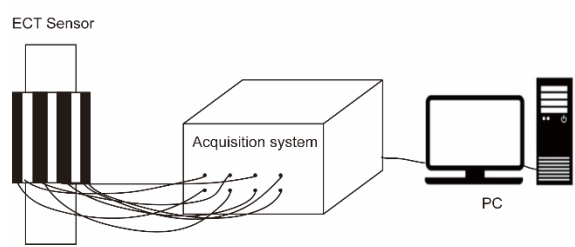

Fig. 6. ERT system

## 4. Results and Discussion

### 4.1 evaluation index

In this paper, we use MSE, SSIM, and PSNR as quantitative criteria for image-reconstruction quality.MSE reflects the error between the generated image and the ground-truth image; essentially, it is the Euclidean distance between the two image matrices. The RMSE is computed as

$$RMSE = \sqrt{\frac{1}{N}\sum_{i=1}^{N}(g_i - \hat{g}_i)^2}$$

where $\hat{g}_i$ is the pixel value produced by the algorithm, $g_i$g\_i$g_i$ is the true pixel value, and $N$N$N$ is the number of pixels (here 256×256).

SSIM (structural similarity). SSIM measures the similarity between two images (or, equivalently, the degree of distortion). Its standard factorized form is

$$SSIM = [I(g - \hat{g})]^{\alpha}[P(g - \hat{g})]^{\beta}[S(g - \hat{g})]^{\gamma}$$

With:

$$l(g_i - \hat{g}_i) = \frac{2\mu_g\mu_{\hat{g}} + m_1}{\mu_g^2 + \mu_{\hat{g}}^2 + m_1}$$

$$P(g_i - \hat{g}_i) = \frac{2\delta_g \delta_{\hat{g}} + m_2}{\delta_g^2 + \delta_{\hat{g}}^2 + m_2}$$

$$S(g - \hat{g}) = \frac{\delta_{g\hat{g}} + m_3}{\delta_g \delta_{\hat{g}} + m_3}$$

Here, $I(g_i - \hat{g}_i)$, $p(g_i - \hat{g}_i)$ and $S(g_i - \hat{g}_i)$ denote the luminance, contrast, and structure comparisons, respectively; μ is the mean, δ the standard deviation, and $\delta_{g\hat{g}}$ the covariance. The exponents α, β, γ and the constants $m_1$, $m_2$, $m_3$ are parameters. In this work, we set α=0 and β=γ=1.

PSNR is a widely used objective image-quality metric that avoids subjective visual judgment. It is defined as:

$$PSNR = 10 \times log_{10}(\frac{peakval}{MSE})$$

where $peakval$ is the maximum possible pixel value.

## 4.2 stimulation results and analysis

We feed the simulated images from the 6,000-sample test set into the generator to obtain GAN-enhanced reconstructions. Figure 7 presents five representative comparisons based on Landweber reconstructions; Figs. 8 and 9 show randomly selected examples based on LBP and Tikhonov reconstructions, respectively. As can be seen, after GAN enhancement the images exhibit clear boundaries with no spurious gray regions; shadows across the field of view are removed and the spatial resolution is markedly improved. We also observe noticeable overfitting beyond 100 training iterations; therefore, we adopt the 100-iteration outputs as the final results.

Figures 10–12 compare quality metrics for 100 samples each of Landweber, LBP, and Tikhonov reconstructions versus their GAN-enhanced counterparts. LBP reconstructions are the most blurred and contain extensive shadowing; Landweber and Tikhonov results show excitation-related shading near boundaries. These artifacts, however, are effectively mitigated by the GAN, indicating that the proposed GAN enhancement benefits all three traditional methods. Overall, the GAN-enhanced images achieve substantially lower RMSE than the originals—by about **22%** on average. While SSIM may decrease slightly for a few cases, the improvement is pronounced for images with initially low SSIM, and the mean SSIM increases modestly. PSNR improvements are particularly large for a subset of images (the affected subset varies by base algorithm); on average, PSNR increases by roughly 9 dB.

Figure 13 summarizes the mean metrics from the preceding figures. The "GAN" values are averaged over the GAN-enhanced outputs of all three baseline algorithms. RMSE and PSNR show substantial gains, and SSIM exhibits a smaller but consistent improvement.

Fig.7 Comparison between Landweber and GAN reconstructions

Fig.8 Comparison between LBP and GAN reconstructions

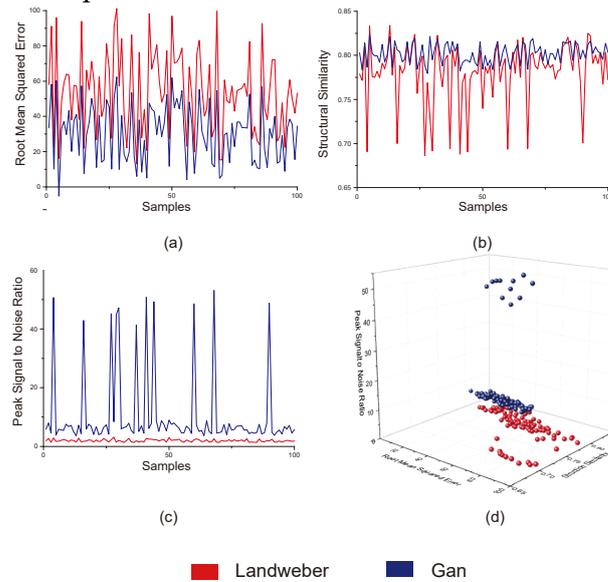

Fig.9 Comparison between Tikhonov and GAN reconstructions

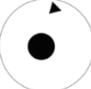

Fig. 10. Comparison between Landweber and GAN reconstructions

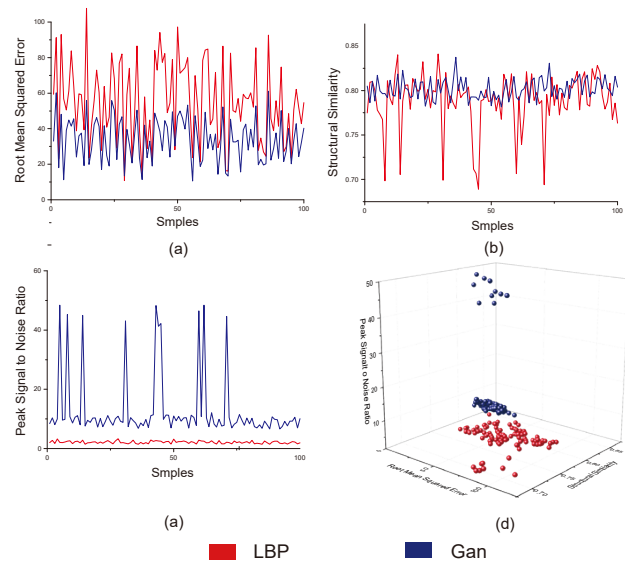

Fig. 11. Comparison between LBP and GAN reconstructions

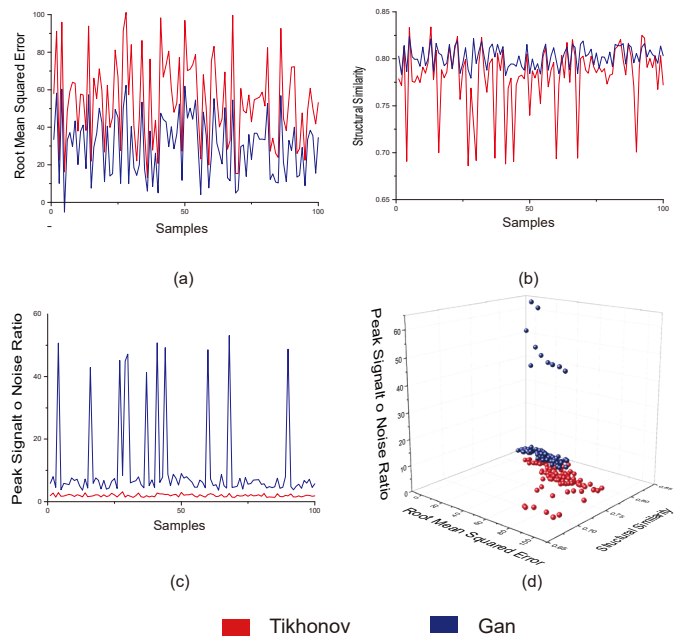

Fig. 12. Comparison between Tikhonov and GAN reconstructions

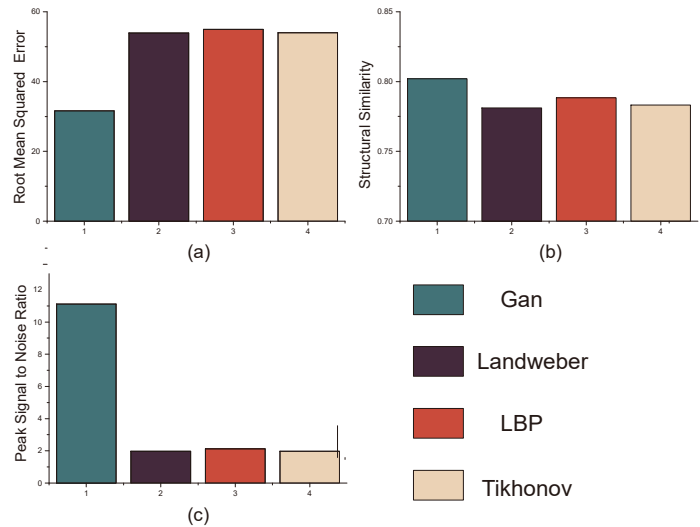

Fig. 13. Mean Comparison between 4 algorithms

## 5. Conclusion

In this paper, we propose a pix2pix-based conditional GAN for ERT image reconstruction. We first generate ERT data via computer simulation, reconstruct images with three conventional algorithms, and then feed those reconstructions into the GAN for refinement. The GAN-enhanced images show significant improvements in objective metrics and, visually, exhibit higher resolution with boundaries that are effectively mesh-independent.

This study has several limitations. (1) Both the simulated and experimental datasets are small, and the inclusions adopt fixed, simple geometric shapes. (2) We only investigate 2D reconstruction; 3D reconstruction is not addressed. Future work will seek to remedy these limitations.

## References

[1] Weifu Fang. A nonlinear image reconstruction algorithm for electrical capacitance tomography. Measurement Science and Technology, 15, 2124.

[2] Karamcheti, Arun, Chakraborti, Nirupam and Kumar Kalra, Prem. Application of Modified Newton-Raphson Methods in Multi-Component Equilibrium Problems.International Journal of Materials Research, 86(1995), 245-252.

[3] Lei Jing, Shi Liu, Li Zhihong, Sun Meng. An image reconstruction algorithm based on the extended Tikhonov regularization method for electrical capacitance tomography. Measurement, 42(2009), 368-376.

[4] Qi Wang, Chengyi Yang, Huaxiang Wang, Ziqiang Cui, Zhentao Gao. Online monitoring of gas–solid two-phase flow using projected CG method in ECT image reconstruction. Particuology, 11(2013), 204-215.

[5] J. Zheng, H. Ma and L. Peng. A CNN-Based Image Reconstruction for Electrical Capacitance Tomography. 2019 IEEE International Conference on

Imaging Systems and Techniques (IST), Abu Dhabi, United Arab Emirates(2019), 1-6.

[6] Anna Fabijańska, Robert Banasiak. Graph convolutional networks for enhanced resolution 3D Electrical Capacitance Tomography image reconstruction. Applied Soft Computing, 110(2021), 107608.

[7] XinJie Wu, MingDa Xu, ChangDi Li, Chong Ju, Qian Zhao, ShiXing Liu. Research on image reconstruction algorithms based on autoencoder neural network of Restricted Boltzmann Machine (RBM). Flow Measurement and Instrumentation, 80(2021), 102009.

[8] Phillip Isola, Jun-Yan Zhu, Tinghui Zhou, Alexei A. Efros. Image-to-Image Translation with Conditional Adversarial Networks. arXiv:1611.07004.

[9]

[10] Andreas Kemna, Jan Vanderborght, Bernd Kulessa, Harry Vereecken. Imaging and characterisation of subsurface solute transport using electrical resistivity tomography (ERT) and equivalent transport models. Journal of Hydrology, 267(2002), 125-146.

[11] Olaf Ronneberger, Philipp Fischer, Thomas Brox. U-Net: Convolutional Networks for Biomedical Image Segmentation. arXiv:1505.04597.

[12] Ugur Demir, Gozde Unal. Patch-Based Image Inpainting with Generative Adversarial Networks. arXiv:1803.07422.

[13] Ronneberger, O., Fischer, P., Brox, T. "U-Net: Convolutional Networks for Biomedical Image Segmentation." *MICCAI 2015*, LNCS 9351, pp. 234–241.